\documentclass[preprintnumbers,nofootinbib,amsmath,amssymb,twocolumn]{revtex4}
\usepackage{amsmath}
\usepackage{amsfonts}
\usepackage{amssymb}
\usepackage{graphicx}    
\usepackage{amsfonts}
\usepackage{amsmath}
\usepackage{braket}
\usepackage{rotating}
\usepackage{verbatim}
\usepackage{multirow}

\newcommand{\fnl}{$f_{\rm NL}$}

\begin{document}

\preprint{Imperial/TP/2013/CC/2} \vskip 0.2in

\title{Non-Gaussian signatures of general inflationary trajectories}

\author{Jonathan~S.~Horner}
\affiliation{Theoretical Physics, Blackett Laboratory, Imperial College, London, SW7 2BZ, UK}
\author{Carlo~R.~Contaldi}
\affiliation{Theoretical Physics, Blackett Laboratory, Imperial College, London, SW7 2BZ, UK}
\affiliation{Canadian Institute of Theoretical Physics, 60 St. George
  Street, Toronto, M5S 3H8, On, Canada}
\date{\today}

\begin{abstract}
  
  We carry out a numerical calculation of the bispectrum in
  generalised trajectories of canonical, single--field inflation. The
  trajectories are generated in the Hamilton-Jacobi (HJ) formalism
  based on Hubble Slow Roll (HSR) parameters. The calculation allows
  generally shape and scale dependent bispectra, or dimensionless
  \fnl, in the out-of-slow-roll regime. The distributions of \fnl\ for
  various shapes and HSR proposals are shown as an example of how this
  procedure can be used within the context of Monte Carlo exploration
  of inflationary trajectories. We also show how allowing
  out-of-slow-roll behaviour can lead to a bispectrum that is
  relatively large for equilateral shapes.
\end{abstract}

\maketitle

\section{Introduction}
The recent results from Planck satellite have confirmed that the
universe is well described by the $\Lambda$CDM
model \cite{Ade:2013ktc,Ade:2013uln}. A cornerstone of this model is
the behaviour of the primordial perturbations to the background
homogeneous model which seed the formation of structure in the
observed universe. The model assumes the perturbations are almost
Gaussian and very close to but \emph{not exactly} independent of
scale. The latter statement following from the observational bounds on
the scalar-spectral index $n_{s} = 0.9603 \pm 0.0073$ \cite{Ade:2013uln}.

A period of accelerated expansion in the very early universe driven by
the potential energy of a slowly evolving scalar field, the inflaton,
 \cite{Starobinsky198099,PhysRevD.23.347,PhysRevLett.48.1220,1982PhLB..108..389L,Linde1983177,mukhanov1981quantum,mukhanov1982vacuum,hawking1982development,guth1982fluctuations,Starobinsky1982175,PhysRevD.28.679,Mukhanov:1985rz}
is the most commonly accepted explanation for the near scale
invariance of the primordial perturbations on scales larger than the
Hubble length. The inflation scenario also explains why the universe
is very homogeneous, isotropic and devoid of monopoles. Inflation
has been criticised on the grounds of requiring fine tuning 
 \cite{Gibbons:2006pa,Hollands:2002xi,Kofman:2002cj,Hollands:2002yb}
and alternatives have been proposed (see e.g. \cite{Brandenberger:2009jq, Brandenberger:2012zb, Creminelli:2010ba, Geshnizjani:2013lza, Avelino:2012ue}),
however none are as simple as the basic inflation scenario involving a
single scalar field.

This statement is simultaneously Inflation's greatest strength and
weakness since the observational bounds on $n_s$ can be satisfied
easily by a large selection of potentials defining even the simplest
single field model. To pin down the exact model of inflation more
precise observations that can constrain higher order statistics of the
perturbations will be required. This is particularly important if even
more complicated models requiring multiple fields are to be
constrained.

A wealth of information could be gained by measuring the
non-Gaussianity of the perturbations. If Inflation did occur then the
deviations from scale-independence and a pure Gaussian distribution
are inherently linked. In the simplest cases both are small and of
order the slow-roll parameter $\epsilon$, representing deviations from
pure de-Sitter space \cite{Stewart:1993bc,Maldacena:2002vr}. Non-Gaussianity is encoded in
the bispectrum, or 3--point function of the perturbations.  The
bispectrum has a much richer structure than the power spectrum as it
is, in principle, a function of three different scales and therefore
contains a lot more information. It may therefore be a very effective
tool for breaking the degeneracy of inflationary models. The
bispectrum is often parametrised by the dimensionless quantity \fnl\
\cite{Komatsu:2001rj}. Most often \fnl\ is quoted in some limit for
the configuration of the mode triangle involved in the 3--point
function and in addition it is usually assumed to be very nearly scale
invariant. Thus \fnl\ is usually regarded as a single amplitude for a
particular configuration of the 3--point function.

The calculation of \fnl\ from inflationary models has received a lot
of attention in recent years \cite{2004PhR...402..103B}. In particular
much focus has been placed on models which generate a large value of
\fnl\ yet retain the near scale invariance of the observed power
spectrum
 \cite{Tzirakis:2008qy,Noller:2011hd,Seery:2005gb,Silverstein:2003hf,Wands:2007bd}. It
was hoped that a large \fnl\ could be observed, potentially confirming
any theory matching the amplitude and shape dependence of \fnl, or at
the very least, ruling out all the models which do not. Unfortunately,
this did not happen with the Planck satellite results which showed
that \fnl\, as measured from Cosmic Microwave Background (CMB)
anisotropies, is consistent with zero with standard deviation of
${\cal O}(10)$ in all ``types'' of \fnl\ \cite{Ade:2013ydc}. This
means the simplest models of inflation are still perfectly consistent
with observations.

Despite this, an accurate calculation of \fnl\ will still be valuable
in future as bounds get stronger and stronger.  This is particularly
important for comparisons with future Large Scale Structure (LSS)
surveys that may constrain \fnl$\sim {\cal O}(1)$ (see e.g. \cite{Dalal:2007cu,Giannantonio:2011ya}). Obtaining {\sl
  accurate} estimates of the bispectrum and its scale dependence for generic
inflationary solutions will be important for these comparisons. This
work will require a numerical evaluation of the primordial bispectrum
arising from higher-order correlations of the curvature
perturbations. The full numerical treatment of the bispectrum has received
little attention over the years, most calculations being analytical
and relying on various approximations. Most numerical work carried out
so far has been concerned only with specific potentials with features
that are known to result in large non-Gaussianity and still rely on
slow-roll approximations to simplify the
calculations \cite{Chen:2006xjb,Chen:2008wn,Hazra:2012yn,Funakoshi:2012ms}.

This {\sl paper} describes the full numerical calculation of
non-Gaussianity for inflationary, single-field trajectories generated
in the Hamilton-Jacobi (HJ) formulation
 \cite{PhysRevD.42.3936}. Initial results from this treatment were
reported in \cite{Horner:2013tma}. The numerical treatment allows the
calculation of non-Gaussianity in cases where the field is not in the
slow-roll regime, but still in the perturbative regime where the
higher-order interaction couplings are still $\ll 1$. It also allows
us to calculate the contribution to all possible ``shapes'' and
``types'' of non-Gaussianity.

In this framework large ensembles of inflating solutions, or {\sl
  trajectories}, can be generated. These are related to a large class
of single field potentials and can, in principle, be compared to
observations without restrictions on the the model of inflation
 \cite{next}. Here we examine the resulting distribution in various
shapes of {\sl local} type non-Gaussianity and verify the well-known
consistency relation for squeezed, single-field inflation
 \cite{Maldacena:2002vr,Creminelli:2004yq}. We also confirm that the
equilateral configuration of the bispectrum follows a similar distribution.

The {\sl paper} is organised as follows. In Section \ref{HJ_traj} we
outline the HJ approach and the analytical framework we are using for
our computations. In Section \ref{comp_method} we describe our
computational method, recapping the calculation of the power spectrum,
followed by the subtleties involved in the calculation of the
bispectrum. In Section \ref{results} we outline the main results of
the paper and verify them through some simple consistency checks. We
discuss our results in Section \ref{conclusion}.

\section{Hamilton Jacobi approach to inflationary trajectories}\label{HJ_traj}
We start by briefly reviewing the HJ approach to inflationary
trajectories where we consider the Hubble-Slow-Roll (HSR) parameters
to be the fundamental quantities of interest, as opposed to the
frequently used Potential-Slow-Roll (PSR) parameters
 \cite{PhysRevD.42.3936,Adshead:2008vn,Liddle:1994dx,Kinney:1997ne}.

If $\phi$ is a monotonic function of time, we can change the
independent variable in
the Friedmann equations from $t$ to $\phi$ and consider all quantities
as functions of $\phi$. The Friedmann equation and the inflaton's
equation of motion then take on the following form
\begin{eqnarray}
\dot{\phi} & = & -2M^{2}_{\rm pl}H'(\phi)\,,\\
  \left[H'(\phi)\right]^{2} - \frac{3}{2M^{2}_{\rm pl}}H(\phi)^{2} & = & -\frac{1}{2M^{4}_{\rm pl}}V(\phi)\,,
  \label{Friedmann}
\end{eqnarray}
where overdots and primes denote a derivative with respect to $t$
and $\phi$ respectively, $H$ is
the Hubble rate, and $M_{\rm pl}^{2} = (8\pi G)^{-1}.$ One of the
advantages of performing this change of variable is that one can
merely pick a function $H(\phi)$ and this will correspond to an exact
solution of a corresponding potential $V(\phi)$. It is straightforward
to verify that inflation will occur if the following condition holds
\begin{equation}\label{epsilon}
\epsilon = 2M^{2}_{\rm pl}\left[\frac{H'(\phi)}{H(\phi)}\right]^{2} \equiv -\frac{\dot{H}}{H^{2}} \equiv \frac{\dot{\phi}^{2}}{2M_{\rm pl}^{2}H^{2}} < 1\,.
\end{equation}
This relation is exact, unlike the equivalent expression for the PSR
parameter $\epsilon_{V} \propto (V'/V)^{2} < 1$ which is only
approximate. 

We can define an infinite hierarchy of HSR parameters labeled by index $l$
\begin{equation}\label{lambda}
^{l}\lambda = \left(2M^{2}_{\rm pl}\right)^{l} \frac{(H')^{l-1}}{H^{l}}\frac{d^{(l+1)}H}{d\phi^{(l+1)}}\,.
\end{equation}
From these we can define $\eta \equiv \,^{1}\lambda =
-(\ddot{\phi}/H\dot{\phi})$ and $\xi \equiv \,^{3}\lambda$. The last
ingredient required is the number of $e$-foldings $N$ specifying the
change in scale factor $a$ during the inflationary phase
$\ln(a)=N$. It is useful to relate this to the Hubble rate as
\begin{equation}\label{efoldings}
\frac{\mathrm{d}N}{\mathrm{d}t} = H\,.
\end{equation}
Combining all of these equations produces the following set of differential equations dictating the evolution of the background
\begin{eqnarray}\label{background}
\frac{\mathrm{d}H}{\mathrm{d}N} & = & -\epsilon H\,,\nonumber\\
\frac{\mathrm{d}\epsilon}{\mathrm{d}N} & = & 2\epsilon(\epsilon - \eta)\,,\\
\frac{\mathrm{d}^{l}\lambda}{\mathrm{d}N} & = & \left(l\epsilon - (l - 1)\eta\right)^{l}\lambda - ^{l+1}\lambda\,.\nonumber
\end{eqnarray}

This is  the most  natural set of  variables to use  when describing a
general inflationary trajectory. These equations will be the starting
point  of our  \fnl\ calculation. The HSR  parameters will
evolve  in time  and  each particular inflation model with a
particular set of initial conditions will  correspond to a distinct
trajectory in  HSR-space. In other words,  specifying the HSR parameters at some
particular  time and solving  the system (\ref{background})  is precisely
equivalent to specifying $\phi(t_{0})$, $\dot{\phi}(t_{0})$, and $V(\phi)$ and
solving the Friedmann equations. 

The HJ system (\ref{background}) is an infinite hierarchy of equations
that describe {\sl all} possible background solutions. For the purpose
of computing observables the system is usually truncated by fixing 
$^l\lambda =0$ for $l\ge l_{\rm max}$. The truncated system still describes {\sl exact}
solutions for the background quantities but restricts the space of
solutions to a subset of the infinite system. 

Relating the HSR picture to a specific model is straightforward for
the simplest cases. For example if we set $^{l}\lambda = 0$ for all $l > 1$ the only
remaining non-zero HSR parameters are $\epsilon$ and $\eta$. This implies
$H(\phi) = a\phi^{2} + b\phi + c$ is a quadratic function and hence
$V(\phi)$ is quartic. If one specifies an initial condition $H_{0}$
this fixes the potential $V(\phi)$ up to a constant shift $\phi \to
\phi + C$. This shift will have no impact on observations because the
energy scale is specified by $H_{0}$. We can use this symmetry to
remove the linear term in $H(\phi)$ and write the potential as
\begin{equation}\label{potential_example}
V(\phi) = \frac{\lambda}{4!}\phi^{4} + \frac{m^{2}}{2}\phi^{2} + \Lambda\,.
\end{equation}
If one specifies $\epsilon_{0}$ and $\eta_{0}$ at the same time as
$H_{0}$ this is then equivalent to solving for the model parameters
and initial conditions
\begin{eqnarray}
\phi_{0} & = & \pm \frac{\sqrt{2\epsilon_{0}}}{\eta_{0}}M_{\rm pl}\,,\nonumber\\
\dot{\phi}_{0} & = & \mp \sqrt{2\epsilon_{0}}H_{0}M_{\rm pl}\,,\nonumber\\
\frac{\lambda}{4!} & = & \frac{3H_{0}^{2}\eta_{0}^{2}}{16 M_{\rm pl}^{2}}\,,\\
\frac{m^{2}}{2} & = & \frac{H_{0}^{2}}{2}(3\eta_{0} - \frac{3}{2}\epsilon_{0} - \eta_{0}^{2})\,,\nonumber\\
\Lambda & = & \frac{2}{27}\lambda M_{\rm pl}^{4}\left(1 + \frac{27}{2}\frac{m^{2}}{\lambda M_{\rm pl}^{2}}\right)^{2}\,.\nonumber
\end{eqnarray}
Note that although we have three degrees of freedom we cannot specify
$\lambda$, $m^{2}$, and $\Lambda$ independently. This is simply
because we have used our freedom in initial condition $\phi_{0}$ to write $H$ as
$H(\phi) = a\phi^{2} + c$. This leaves two degrees of freedom to
specify $\lambda$, $m^{2}$ and $\Lambda$. In practice, if one only
requires the shape of the potential $V(\phi)$ it is much simpler to
solve for $\phi(N)$, $H(N)$, and
$\epsilon(N)$ and use the relation
\begin{equation}\label{potential}
V(\phi) = 3M^{2}_{\rm pl}H^{2}\left(1 - \frac{\epsilon}{3}\right)\,.
\end{equation}

The only remaining information that needs to be specified in the model
above is the total number of $e$-foldings $\Delta N$. When integrating
the Friedmann equations for a given potential $V(\phi)$ there is no
clear way of ensuring inflation ends, or if it provides enough
inflation. Inflation ends exactly when $\epsilon = 1$. The only
constraint on the length of inflation is that it must last at least
roughly 60 $e$-foldings \cite{Ade:2013uln} in order for all scales up
to the present Hubble scale to have been inflated to super horizon
scales before the deceleration phase of the standard Big Bang
picture. Converting this into some length in time necessarily requires
some knowledge of $H$ (which may vary significantly over the whole of
inflation) so $N$ is clearly the most natural time variable to
use. These constraints on inflation are then easy to implement using
the HSR parameter system - to ensure inflation ends we choose the
initial condition $\epsilon(N_{\text{tot}}) = 1$. To ensure inflation
provides enough $e$-foldings we integrate {\sl back} in time from
$N_{\text{tot}} \to N = 0$ where $N_{\text{tot}} \sim 60$. In practice
the exact value of $N_{\text{tot}}$ is not known due to uncertainties
in the physics of reheating. When generating random trajectories
$\Delta N$ can be drawn from a proposal density distribution to account for this
uncertainty.

To generate large ensembles of random inflationary trajectories we can
then draw the remaining HSR parameters $^{l}\lambda$ at the end of inflation from
proposal densities. In the following the proposal densities are uniform
over a specified range in each HSR but could also take different forms
e.g. normal distribution. The choice of proposal shape and {\sl where} the
boundary conditions are drawn can lead to significant differences in
the distributions of the final observable quantities. A number of
different choices have been made in the literature
 \cite{Kinney:1997ne,Easther:2002rw,2005PhRvD..72h3520C,Liddle:2003py}.

It is important to emphasise that the evolution of these trajectories
need not have anything to do with inflationary dynamics as $H(\phi)$ can
be completely decoupled from the system. One is perfectly able to
solve for $\epsilon(N)$, $\eta(N)\dots$ without mentioning
inflation. The key ingredients to connect with inflation are $H(N)$
and $V(\phi)$ or (\ref{potential}), both of which only require
an input function $\epsilon(N)$. The HSR parameters themselves, along
with their differential equations, only provide an efficient tool for
generating valid functions $\epsilon(N)$ which may {\sl then} be
correctly interpreted as inflationary models \cite{Liddle:2003py}.

\subsection{Monte Carlo generation of HJ trajectories}

The generation of large ensembles of consistent inflationary trajectories
in the HJ formalism lends itself to Monte Carlo Markov Chain (MCMC)
comparisons of the inflationary model space with observations such as
the Planck CMB measurements. The HSR definition is
particularly useful since in the slow roll limit the proposal
parameters are closely related to the observables such as $n_s$, the
tensor-to-scalar ratio $r$, running $dn_s/d\ln k$ etc. For example, at
second order in HSR parameters
\begin{eqnarray}
n_{s} & = & 1 - 4\,\epsilon + 2\,\eta - 2\,(1 + C)\,\epsilon^{2} -
\nonumber\\
&&\frac{1}{2}(3 - 5C)\,\epsilon\,\eta + \frac{1}{2}(3 - C)\,\xi\,,\label{nsr_eqn1}\\
r & = & 16\,\epsilon\,[1 + 2C(\epsilon - \eta)]\,,\label{nsr_eqn2}\\
n_t &=& -2\, \epsilon + (3+C)\,\epsilon^2+(1+C)\, \epsilon\,\eta\label{nsr_eqn3}\,,
\end{eqnarray}
where $C = 4(\ln 2 + \gamma) - 5$ and $\gamma$ is the Euler-Mascheroni
constant \cite{Stewart:1993bc}. As described below we calculate all
observables numerically and use (\ref{nsr_eqn1})-(\ref{nsr_eqn3}) for comparison.

Here, we explore the proposal densities for observables resulting from
the HJ formalism and including non-Gaussianity. The use of the
proposal densities for comparison with the data will be explored in
 \cite{next}. As a simple assumption for the proposal densities from
which to draw HSR boundary conditions we use uniform distributions in
the range
\begin{equation}\label{uniform}
^{l}\lambda = [-1,1]e^{-s\,l}\,,
\end{equation}
for $l>0$ and where $s$ is a suitable suppression factor. Our boundary
conditions will be imposed at the end of inflation so $\epsilon=1$ and
$N_{\text{tot}}$ is also drawn from a uniform distribution
$N_{\text{tot}} = [60,80]$. In our formulation $N$ increases with time
so $N \sim 0$ represents the time at which the largest scales
observable today were exiting the horizon and $N=N_{\text{tot}}$ is
the end of inflation. The observable window spanned by e.g. CMB
observations corresponds approximately to the interval $N\sim 0
\rightarrow N\sim 10$. Note that the normalisation of $H$ does not
affect the evolution of the parameters so we may specify the initial
condition for $H$ at any time in order to correctly normalise the
amplitude of perturbations. In practice we have to truncate the HSR
series for some finite value of $l = l_{\text{max}} - 1$ (so
$l_{\text{max}} = 3$ implies $\epsilon$, $\eta$, $\xi$ are non-zero)
\footnote{An alternative ``model-independent'' method is to
  parametrise the potential via a Taylor expansion of a certain order
  as done in \cite{Ade:2013uln}. The two method are complementary.}.

\begin{figure}[t]
  \begin{center}
    \includegraphics[width=8.5cm,trim=0cm 0cm 0cm 0cm,clip]{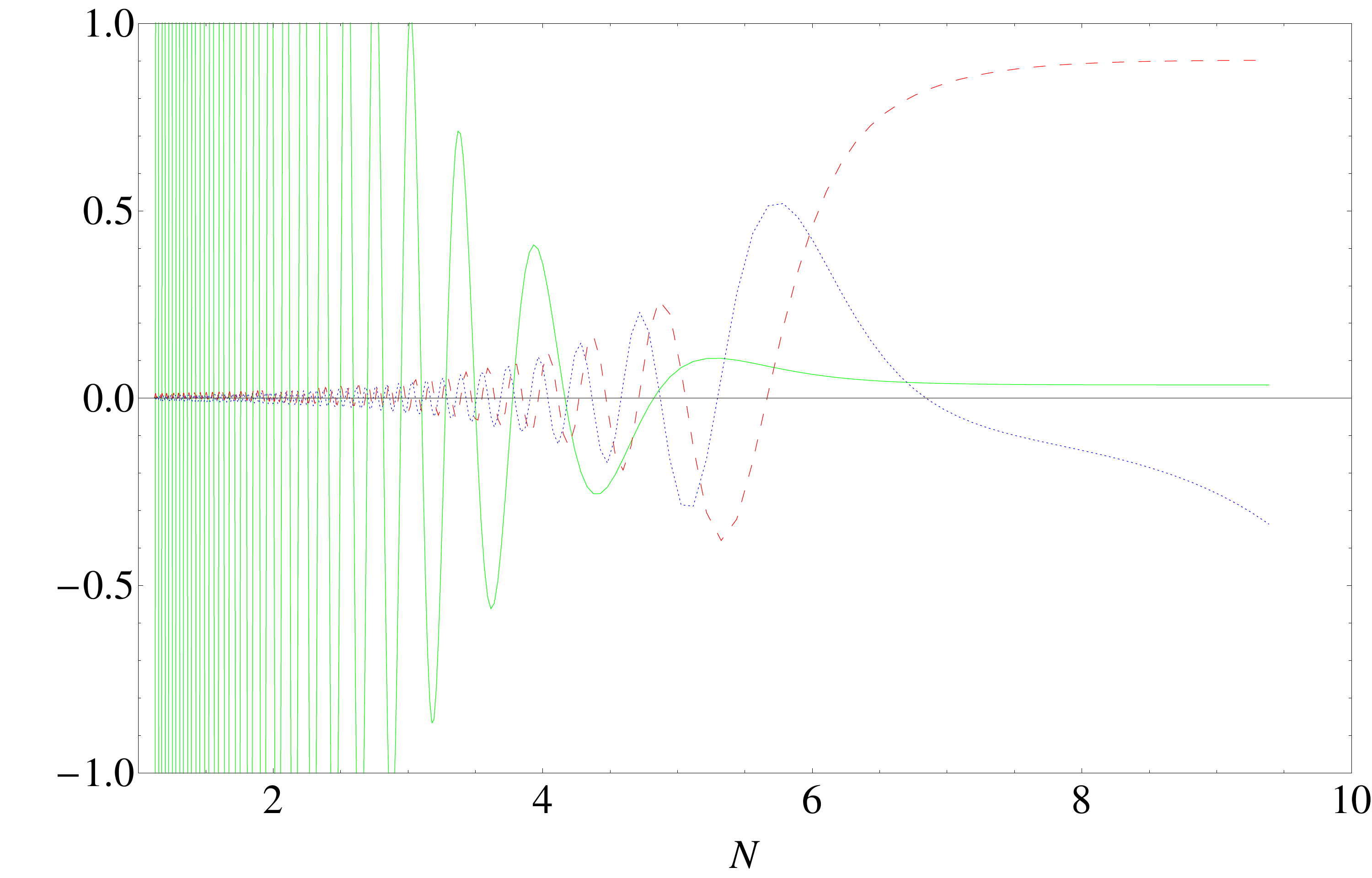} 
    \caption{The evolution of $\zeta$ and \fnl\ as a function of
      $e$-fold $N$ for a typical random
      trajectory. The curves are normalised arbitrarily for the
      purpose of visualisation. The green (solid) line shows the real
      part of $\zeta$ for a  mode that crosses the horizon at $N\sim
      6$. $\zeta$ converges to a constant shortly after horizon
      crossing as expected. The blue (dotted) and red (dashed)
      curves show the evolution of the real and imaginary parts of the
      integral in (\ref{fnl_final}). Only the
      imaginary part that converges after horizon exit contributes to
      the value of \fnl\ whilst the real, diverging component is discarded.}
    \label{fig:time_dependance}
     \end{center}
\end{figure}

\begin{figure}[t]
  \begin{center}
    \begin{tabular}{c}
     \makebox[8.5cm][c]{
    \includegraphics[width=8.5cm,trim=0cm 0cm 0cm
    0cm,clip]{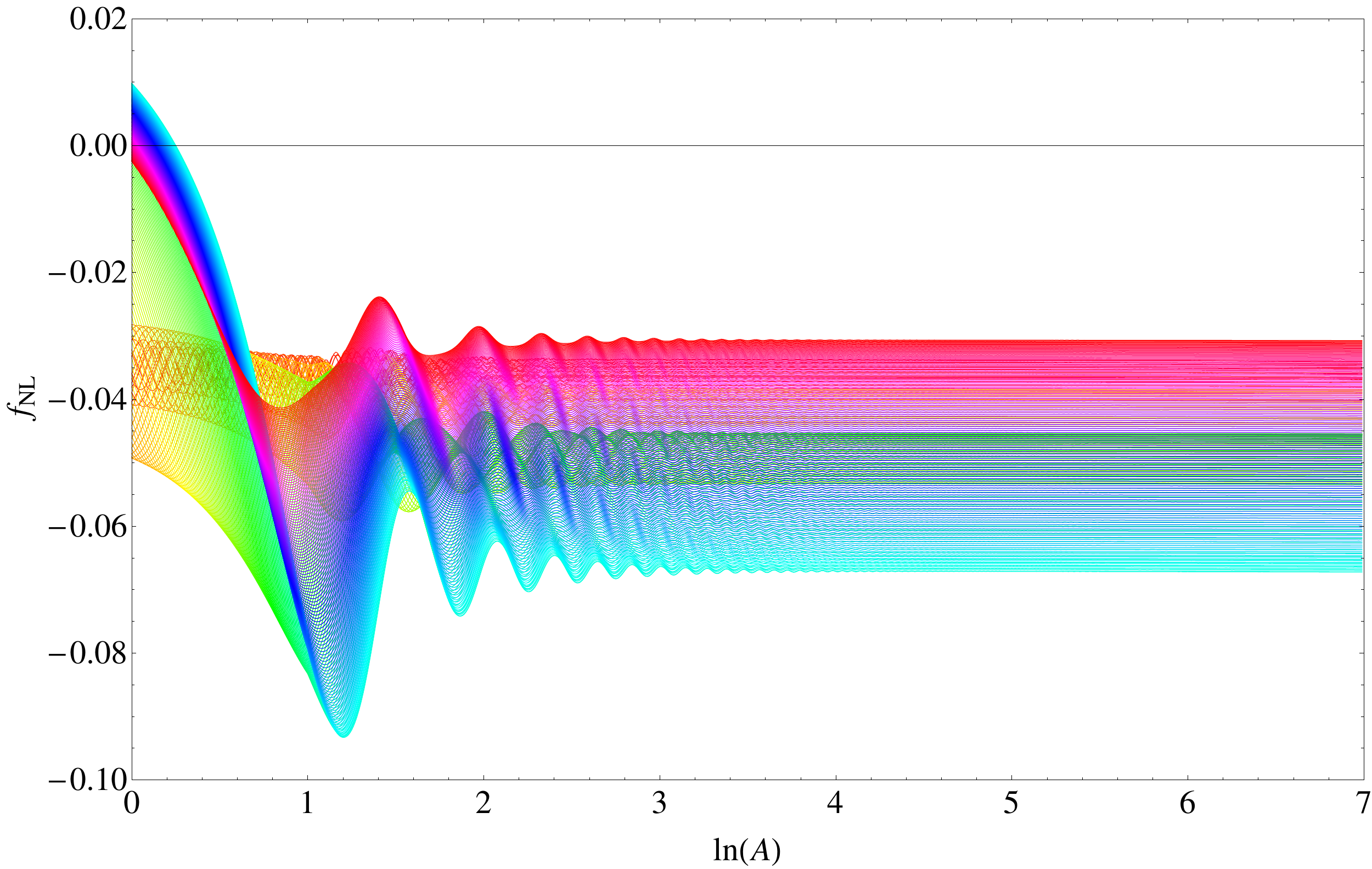}}\\
  \makebox[8.5cm][c]{
    \includegraphics[width=8.5cm,trim=0cm 0cm 0cm
    0cm,clip]{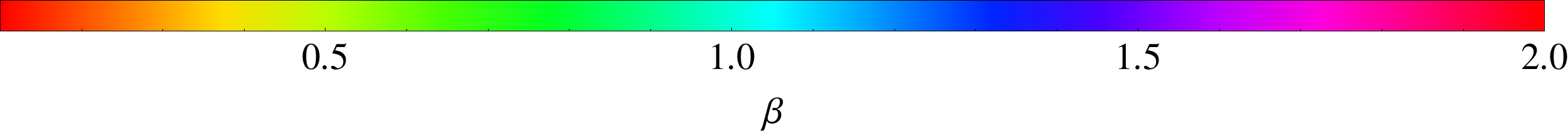}}
    \end{tabular}
    \caption{This figure shows how \fnl\ for different shape parameter
      $\beta$ depends on the integration start scale parameter
      $A$. Each of the curves is generated from the same HSR
      trajectory for comparison. The parameter $A$ represents how deep
      inside the horizon the mode {\sl smallest} $k$ in the triangle was at the start
      of the integration. \fnl\ converges for all shapes as $A$
      becomes large, signifying earlier start times with respect to
      horizon exit. Note that, as expected, \fnl\ peaks at roughly
      $\beta \sim 1$. Typically when $A \sim 400$ \fnl\ has converged
      with only residual numerical noise at the a level of $\lesssim
      1$\%. The source of the residual noise is the early-time
      oscillatory integral approximation (see below).}
    \label{fig:early_time}
  \end{center}
\end{figure}

\section{Computational method}\label{comp_method}
\subsection{Computation of the power spectrum}
We introduce a comoving curvature perturbation $\zeta(t,\mathbf{x})$
and work in a gauge where the spatial part of the perturbed metric is
given by $g_{ij} = a^{2}\,(t)e^{2\zeta(t, \mathbf{x})}\delta_{ij}$ and
the inflaton perturbation vanishes everywhere
$\delta\phi(t, \mathbf{x}) = 0$. The primordial power spectrum of the
curvature perturbations is related to the variance of the Fourier
expanded mode $\zeta_k$
\begin{equation}\label{PowerSpectrum}
\langle\zeta^{}_{k_{1}}\zeta^\star_{k_{2}}\rangle = (2\pi)^{3}\delta^{(3)}(\mathbf{k}_{1}+\mathbf{k}_{2})P_\zeta(k_1)\,,
\end{equation}
where $\mathbf{k}$ is the Fourier wavevector and $k\equiv
|\mathbf{k}|$. The mode $\zeta_{k}(t)$ satisfies the Mukhanov-Sasaki
equation \cite{Mukhanov:1985rz,Sasaki:1986hm}. Expressed in terms of
$N$ instead of $t$ this equation becomes
\begin{equation}\label{Mukh}
  \frac{\mathrm{d}^{2}\zeta_{k}}{\mathrm{d}N^{2}} + (3 + \epsilon - 2\eta)\frac{\mathrm{d}\zeta_{k}}{\mathrm{d}N} + \frac{k^{2}}{a^{2}H^{2}}\zeta_{k} = 0\,.
\end{equation} 
In this form it is trivial to see that outside the horizon the
derivative of $\zeta_{k}$ decays exponentially with respect to $N$ or
as $a^{-2}$ so
$\zeta_{k}$ quickly goes to a constant. The power spectrum of interest
is then related to the freeze-out value of $\zeta_k$ on scales $k\ll
aH$
\begin{equation}
  P_\zeta(k) = |\zeta_{k\ll aH}|^{2}\,.
\end{equation}
The initial conditions for the solutions to (\ref{Mukh}) can be set
when the mode is much smaller than the horizon $k\gg aH$ and takes on
the  Bunch-Davies  form \cite{Bunch:1978yq}
\begin{equation}\label{Initial_zeta}
\zeta_{k} \to \frac{1}{M_{\rm pl}}\,\frac{e^{-ik\tau}}{2a\sqrt{k\epsilon}}\,,
\end{equation}
where $\tau$ is conformal time defined by $\mathrm{d}N/\mathrm{d}\tau =
aH$. From (\ref{Mukh}) the phase of $\zeta_{k}$ is irrelevant
and we only need its rate of change for the initial condition on
$\mathrm{d}\zeta_{k}/\mathrm{d}N$ so we never need to explicitly
evaluate $\tau$.

\begin{figure*}[t]
  \begin{center}
    \begin{tabular}{cc}
     \makebox[8.5cm][c]{
    \includegraphics[width=8.5cm,trim=0cm 0cm 0cm
    0cm,clip]{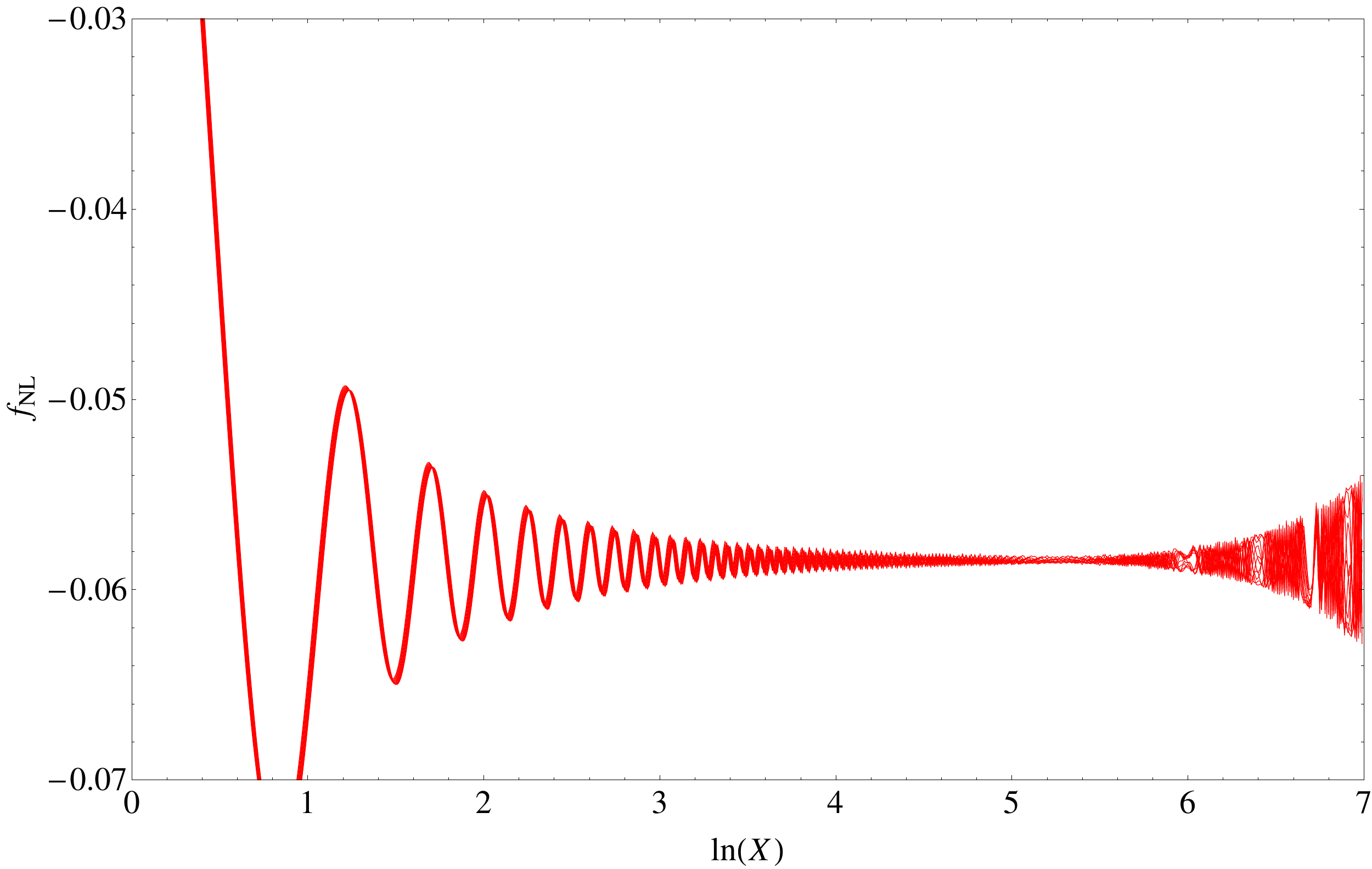}}&
  \makebox[8.5cm][c]{\includegraphics[width=8.5cm,trim=0cm 0cm 0cm
    0cm,clip]{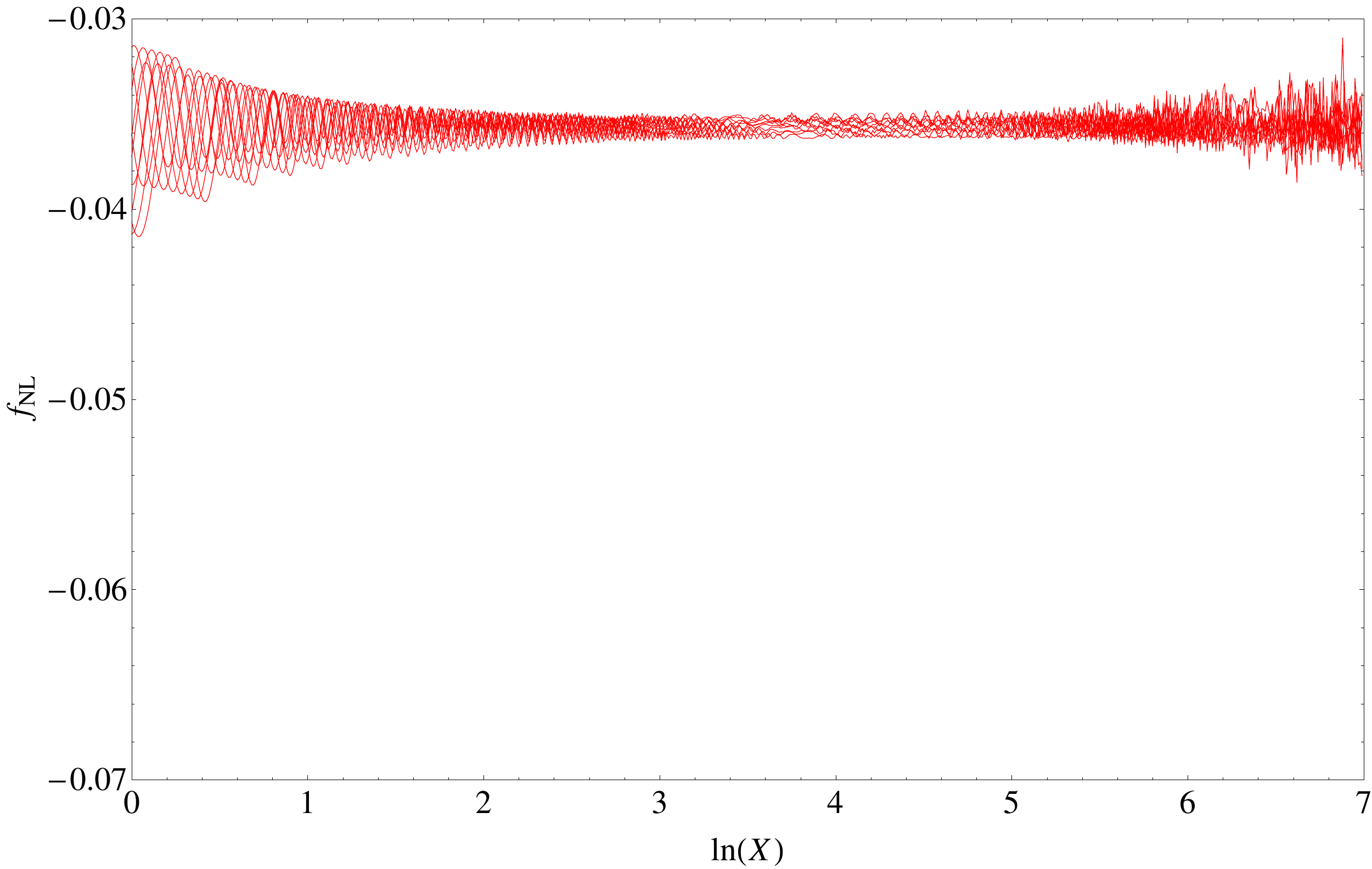} }
  \end{tabular}
  \caption{Left Panel: Dependence of \fnl\ on the position of the
    integral split point parameter $X$. The ten lines are for \fnl\
    from 10 ``equilateral'' shape configurations ($\beta=[0.95,1.05]$)
    for the same HSR trajectory. If the split point is too late,
    $X=k/aH\to 1$ then the WKB approximation used to calculate the
    early contribution from the diverging, oscillating integrand
    breaks down. If the split point is too early then inaccuracies in
    the numerical integration of the oscillatory function start to
    dominate. The optimal value of the split point is found to be $\ln
    X=4\to 5$ where the total noise is $\ll 1$\%. Right panel: same
    but for the ten most ``squeezed'' triangles (i.e. with
    $\beta=[0.1-0.2]$). The optimal value for $X$ is slightly lower in
    this case but still small for the choice $\ln X=4\to 5$.}
    \label{fig:cutoff_point}
  \end{center}
\end{figure*}

For our \fnl\ calculation we are interested in solving this equation
for an observable range of $10^{-5} < k < 10^{-1}$ in units of
$(\text{Mpc})^{-1}$ for each inflationary trajectory obtained via the
HJ system. Each background model is completely
defined from the solutions of (\ref{background}) up to an
overall normalisation of $H$. To choose this normalisation we need to
look at our calculation of $\zeta_{k}$ more closely.

We integrate (\ref{Mukh}) from a time satisfying $k = A\,aH$ to
$k = B\,aH$ where $A \gg 1$ and $B \ll 1$ representing sub and
super-horizon times respectively. Whatever units we wish to work in,
we can fix the normalisation of $a$ so that at $N = 0$ the following
condition is satisfied
\begin{equation}
k_{\text{min}} = A\,aH\,.
\end{equation}
Here $k_{\text{min}}$ represents the smallest $k$ of interest, in
practice the mode corresponding to the largest scales observable
today. For this particular mode one can then approximate the time of
horizon crossing as $N_{c} \approx \ln A$ (this is exact if $H$ is
exactly constant and is the only time we use this approximation). The
initial condition on $H$ will have a direct effect on the amplitude of
the power spectrum. Therefore during the background integration of the
flow parameters we fix the initial condition on $H$ to be
\begin{equation}\label{H_normalisation}
H(N_{c}) = 4\pi \sqrt{2\pi\epsilon(N_{c})}M_{\rm pl}A_{s}\,,
\end{equation}
where $A_{s}$ is the normalisation of the canonical form of the dimensionless
primordial curvature perturbation
\begin{equation}
  k^3\,P_\zeta(k) = A_s\, k^{n_s-1}\,,
\end{equation}
and is typically of the order of $10^{-5}$ to reproduce typical
density fluctuations amplitudes.

\begin{figure*}[t]
  \begin{center}
    \begin{tabular}{cc}
      \makebox[8.5cm][c]{
        \includegraphics[width=6.5cm,trim=0cm 0cm 0cm
        0cm,clip,angle=270]{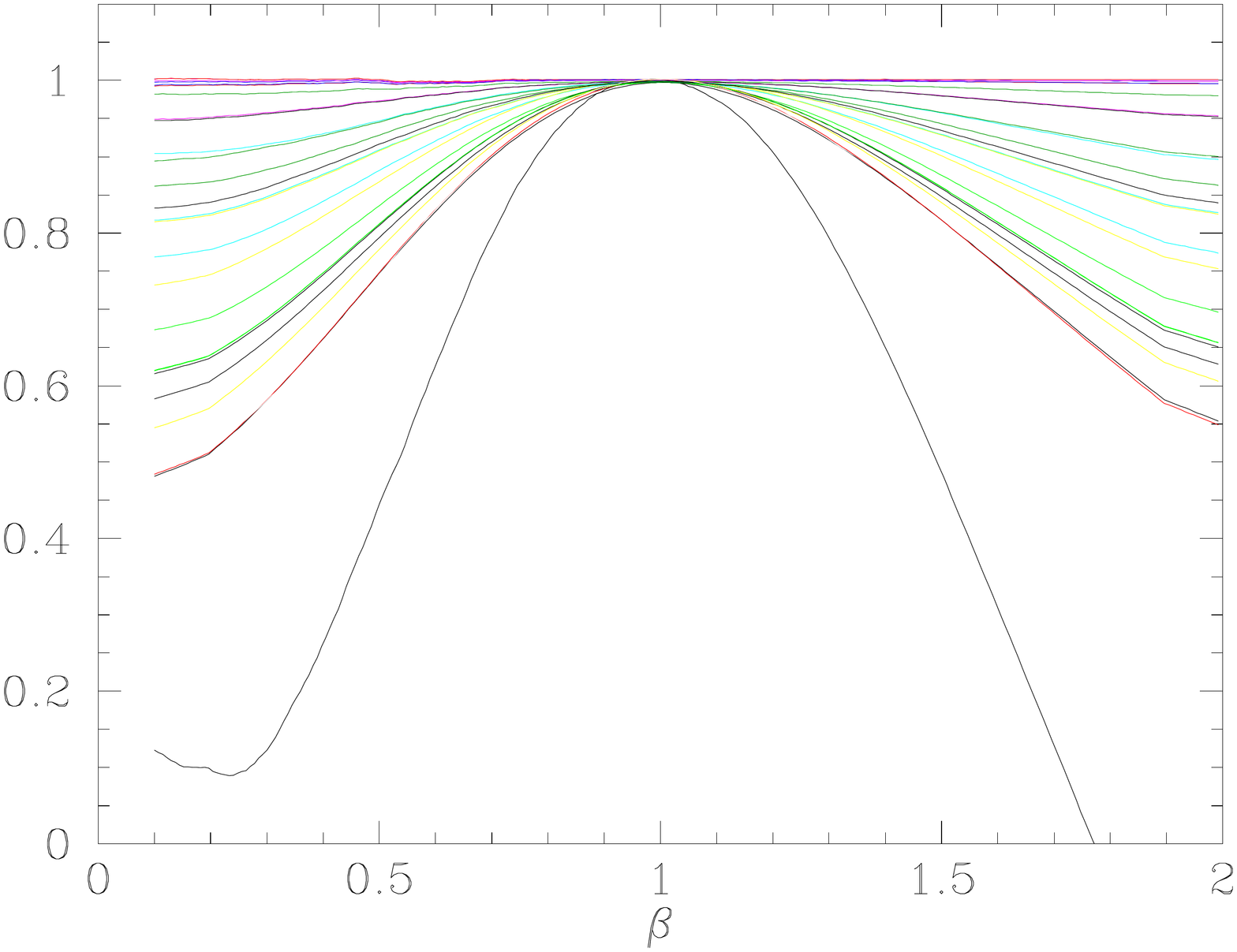}}&
      \makebox[8.5cm][c]{\includegraphics[width=6.5cm,trim=0cm 0cm 0cm
        0cm,clip,angle=270]{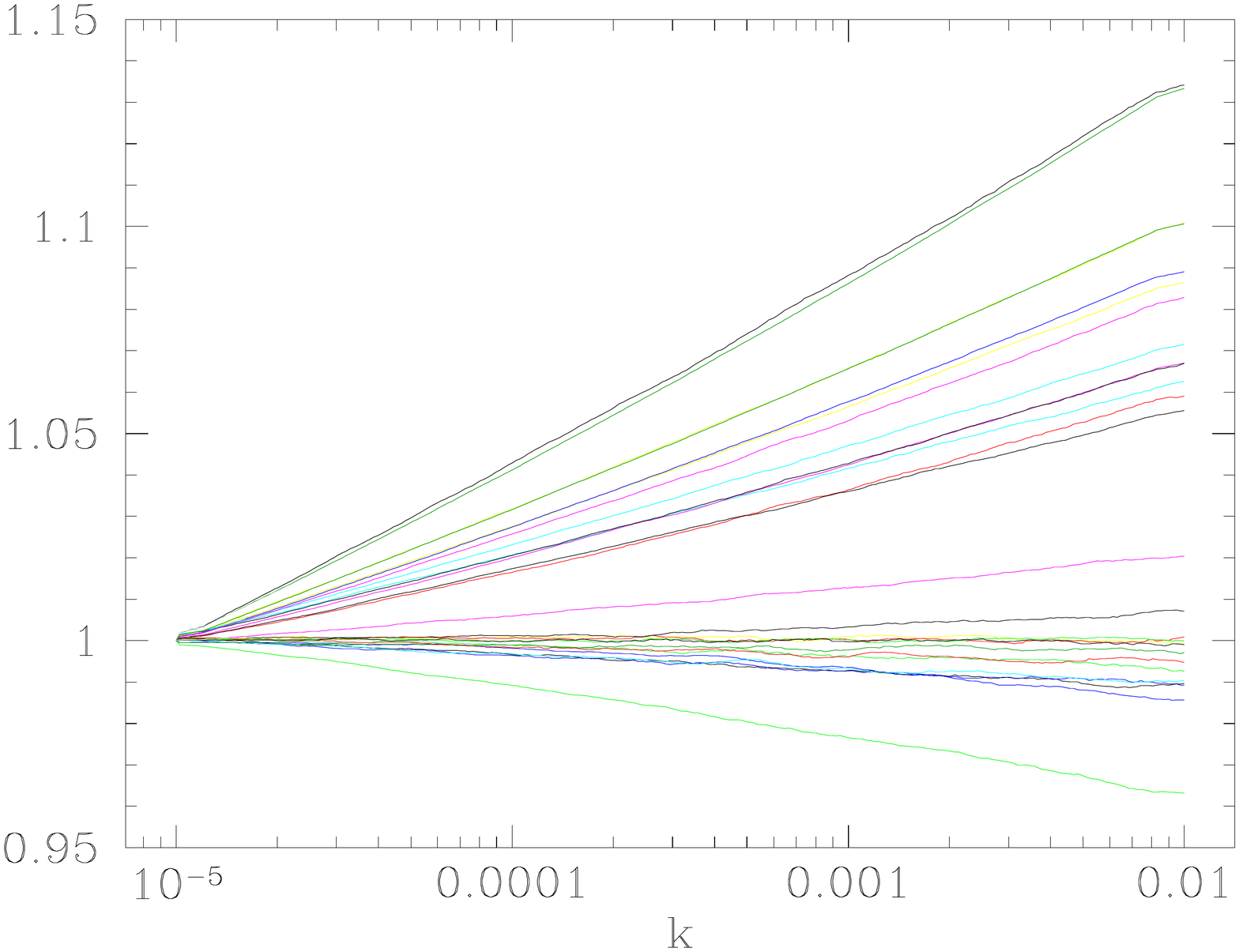}}
    \end{tabular}
    \caption{Shape (left) and scale (right) dependence of \fnl\ for a
      selection of trajectories from the ``end-of-inflation'' boundary
      condition ensemble. The curves have been normalised with respect
      to their value at $\beta=1$ and $k_\star = 10^{-5}
      (\text{Mpc})^{-1}$ respectively.}
    \label{fig:shape_scale_dependence}
  \end{center}
\end{figure*}

We also need to increase the total number of $e-$folds $N_{\text{tot}}
\to N_{\text{tot}} + \ln A$. If this was ignored, as $A$ increases the
mode would start deeper inside the horizon but the initial conditions
on the HSR parameters would remain constant. This would effectively
change the trajectory so the HSR values at horizon crossing would be
different. Shifting the total $e-$folds by $\ln A$ and enforcing
(\ref{H_normalisation}) ensures that $H$ and the HSR
parameters, evaluated at horizon crossing, are independent of $A$ (how
deep the modes start inside the horizon). Neglecting these effects
would affect the convergence of the power spectrum as $A \to \infty$.

A simpler way of normalising $H$ would be to specify the initial
condition at the end of inflation (with all the other HSR parameters)
but that choice is not as physically transparent. In addition, $H$ may
vary by orders of magnitudes during the approximately 60 $e$-foldings
of evolution.  This can lead to a large variation in the overall
normalisation of the primordial power which can lead to numerical
problems if one wishes to use the results as the input to standard
boltzmann codes such as {\tt CAMB} \cite{Lewis:2002ah}.

To be consistent we require (\ref{Initial_zeta}) to be satisfied for
each $k$. Therefore in order for each mode to start ``equally deep''
inside the horizon we integrate the background forward in time (from
$N = 0$) until $k = A\,aH$ for every mode of interest. Applying
(\ref{Initial_zeta}) we integrate the background and (\ref{Mukh})
until each mode crosses the horizon and satisfies $k = B\,aH$. This
ensures the modes have sufficiently converged to their super-horizon
values. In practice it was found that, for the calculation of the
bispectrum, the solutions converged for $A \approx e^{6}$ and $B
\approx 0.1$. Larger values of $A$ significantly added to
computational time due to the erratic early time behaviour of
$\zeta_{k}$ with no real benefit.

This completely determines the mode evolution and hence their value on
super-horizon scales.  We can then calculate physical observables such
as $n_{s}$ and $r$ from their definitions directly without resorting
to any approximations
\begin{eqnarray}\label{nsr_numeric}
n_{s}(k_\star) & = & 1 + \frac{\mathrm{d}\ln \left[k^{3}P_{\zeta}(k_\star)\right]}{\mathrm{d}\ln k}\,\\
r(k_\star) & = & 2\,\frac{P_h(k_\star)}{P_{\zeta}(k_\star)}\,\nonumber
\end{eqnarray}
where we evaluate the quantities at a scale $k_\star$ normally chosen
to be the largest mode in the system.  $P_h$ is the power spectrum of
either the tensor mode polarisations $h_+$ and $h_\times$. The factor
of 2 accounts for the fact that in parity invariant models both
polarisations contribute the same exact power. Solutions for both
gravitational wave polarisations can be obtained by integrating an
equation similar to (\ref{Mukh})
\begin{equation}\label{tensors}
\frac{\mathrm{d}^{2}h_{k}}{\mathrm{d}N^{2}} + (3 - \epsilon)\frac{\mathrm{d}h_{k}}{\mathrm{d}N} + \frac{k^{2}}{a^{2}H^{2}}h_{k} = 0\,,
\end{equation} 
with initial condition
\begin{equation}
h_{k} \to \frac{1}{M_{\rm pl}}\,\frac{e^{-ik\tau}}{a\sqrt{2k}}\,,
\end{equation}
in the limit where $k \gg aH$.

It is worth noting that choosing $B = 1$ (terminating \emph{exactly}
at horizon crossing) produces the the best agreement between equations
(\ref{nsr_numeric}) and (\ref{nsr_eqn1})-(\ref{nsr_eqn3}) and for very small values of $B$
the results can disagree by $\mathcal{O}(\epsilon)$. This is purely
because the slow-roll parameters \emph{evolve} while the power
spectrum remains constant and so the slow-roll formula (which is
specified at horizon-crossing) ceases to be valid for sufficiently
small $B$. This gives us confidence in our numerical results.

It is important to stress that our choice of priors (in particular our
choice of \emph{location} for the priors) typically generates
trajectories where the HSR parameters become small during the time we
calculate $P_{k}$. But the method outlined above works for
\emph{arbitrary} values of these parameters. We could specify the
initial conditions at the beginning of inflation to begin with, easily
breaking slow roll, but we cannot guarantee the trajectory will
provide enough inflation.

\subsection{Computation of the bispectrum}
The non-Gaussianity of the primordial curvature perturbations is
encoded in the third order moment of $\zeta_k$ which, in the isotropic
limit, is a function of the wavenumbers of three wavevectors forming
closed triangles in momentum space
\begin{equation}\label{3rd}
\langle\zeta_{k_{1}}\zeta_{k_{2}}\zeta_{k_{3}}\rangle  =  (2\pi)^{3}\delta^{(3)}(\mathbf{k}_{1}+\mathbf{k}_{2}+\mathbf{k}_{3})B(k_{1}, k_{2}, k_{3})\,.
\end{equation}
For convenience the bispectrum $B$ is re-written in a dimensionless form
$f_{\rm NL}(k_{1}, k_{2}, k_{3})$
by dividing it by different combinations of the squares of the power
spectra of the three modes. \fnl\ is defined in terms of the bispectrum
\cite{Komatsu:2001rj}
\begin{eqnarray}\label{fnl_def}
  f_{\mathrm{NL}}(k_{1}, k_{2}, k_{3})  &=&  \frac{5}{6}\,B(k_{1}, k_{2},
  k_{3}) / \left(|\zeta_{k_{1}}|^{2}|\zeta_{k_{2}}|^{2}+\right.\nonumber\\
  &&\left.|\zeta_{k_{1}}|^{2}|\zeta_{k_{3}}|^{2}+|\zeta_{k_{2}}|^{2}|\zeta_{k_{3}}|^{2}\right)\,,
\end{eqnarray}
and the $5/6$ factor has been introduced by convention.

The weighting introduced in (\ref{fnl_def}) is often called the
``local'' type and others have also been used when motivated by the
expected signal-to-noise of different shaped triangles in the
observations. In particular \cite{Ade:2013ydc} analysed the data with
respect to two additional weightings - equilateral and orthogonal. The
limits reported in \cite{Ade:2013ydc} are $f_{\rm NL}^{\rm local} = 2.7
\pm 5.8$ , $f_{\rm NL}^{\rm equil} = -42\pm 75$, $f_{\rm NL}^{\rm
  ortho}= -25\pm 39$.

The \fnl\ function is normally reduced to a single, scale invariant
amplitude for a particular shaped triangle, as above. This motivates the
different choice of weightings in analysing observations and reporting
results. In our case we will consider the $k_1$, $k_2$, $k_3$
dependence of \fnl\ explicitly and the choice of weighting in relating
the bispectrum to the dimensionless \fnl\ is irrelevant. Throughout
this work we use (\ref{fnl_def}) as the definition of \fnl\ even when
we take the limit of different shaped triangles.

In order to calculate \fnl\ the third order correlator of (\ref{3rd})
needs to be calculated at late times in the super-horizon limit. To do
this we consider the expansion of the action for $\zeta$ at third
order which in terms of the HSR parameters can be written as
 \cite{Maldacena:2002vr, Noller:2011hd, Seery:2005gb}
\begin{widetext}
\begin{eqnarray*}\label{action}
  S_{3} &=& M^{2}_{\rm pl} \!\!\int d^4x\, \left[ a^{3}\epsilon^{2}\zeta \dot{\zeta}^{2} +a\epsilon^{2}\zeta(\partial\zeta)^{2}\right.\nonumber\\
  & &\!\!\!\!\left.  - 2a^{3}\epsilon^{2}\left(1 - \frac{\epsilon}{4}\right)\dot{\zeta}\partial_{i}\zeta\partial_{i}\partial^{-2}\dot{\zeta} + \frac{a^{3}\epsilon^{3}}{4}\partial^{2}\zeta\partial_{i}\partial^{-2}\dot{\zeta}\partial_{i}\partial^{-2}\dot{\zeta} + a^{3}\epsilon\frac{\mathrm{d}}{\mathrm{d}t}\left(\epsilon - \eta\right)\dot{\zeta}\zeta^{2} + 2f(\zeta)\frac{\delta L}{\delta \zeta}\right]\,,
\end{eqnarray*}
\end{widetext}
where $\partial_i\equiv \partial/\partial x_i$, $\partial^2$ and
$\partial^{-2}$ are the Laplacian and inverse Laplacian operators
respectively, and $\delta L/\delta \zeta$ is the equation of motion (\ref{Mukh})
\begin{equation}\label{EoM}
\frac{\delta L}{\delta \zeta} = a\left(\frac{\mathrm{d}}{\mathrm{d}t}\left(a^{2}\epsilon\dot{\zeta}\right) + Ha^{2}\epsilon\dot{\zeta} - \epsilon\partial^{2}\zeta\right)\,.
\end{equation}
The function $f(\zeta)$ is
\begin{eqnarray}\label{f}
f(\zeta) &=& \frac{\epsilon - \eta}{2}\zeta^{2} +
\frac{1}{H}\zeta\dot{\zeta} +\nonumber\\
&&\!\!\!\!\!\!\!\!\!\!\!\!\frac{1}{4a^{2}H^{2}}\left(-(\partial\zeta)^{2}
  - \partial^{-2}\left(\partial_{i}\partial_{j}(\partial_{i}\zeta\partial_{j}\zeta)\right)\right)+\nonumber\\
&& \!\!\!\!\!\!\!\!\!\!\!\!\frac{\epsilon}{2H}\left(\partial\zeta\partial\partial^{2}\dot{\zeta} - \partial^{-2}\left(\partial_{i}\partial_{j}(\partial_{i}\zeta\partial_{j}\partial^{-2}\dot{\zeta})\right)\right)\,,
\end{eqnarray}
which gathers terms proportional to the equation of motion $\delta
L/\delta \zeta$ that do not contribute to the third order action.

In analytical estimates of \fnl\ it is helpful to introduce a number
of field redefinitions that simplify the calculations by suppressing
the terms proportional to $\delta
L/\delta \zeta$ explicitly and isolate the dominant contributions to (\ref{action}) \cite{Maldacena:2002vr, Seery:2005gb}. The redefinitions are not strictly
required when calculating the contributions numerically and introduce
slow-roll approximations which are against the approach being taken
here. The approach described below is equivalent but avoids making some
assumptions inherent in the slow-roll limit.

We are interested in calculating the bispectrum using the ``in-in''
formalism. At tree-level this requires the calculation of \cite{Adshead:2009cb,Maldacena:2002vr,Seery:2005gb}
\begin{equation}\label{in_in}
\langle\zeta^{3}(t)\rangle = -i\int_{-\infty}^{t}\mathrm{d}t'\langle\left[\zeta^{3}(t), H_{\text{int}}(t')\right]\rangle\,,
\end{equation}
where $H_{\text{int}}$, the interaction Hamiltonian, is essentially
$S_{3}$ without the integral over time. Each of the terms in $S_{3}$
contribute separately to the correlation (\ref{in_in}) and can be
considered individually. We are treating $\zeta$ as a quantised curvature perturbation
that is expanded in term of a time dependent amplitude and standard
momentum space creation and annihilation operators
\begin{equation}\label{zeta_quantised}
\zeta(t,\mathbf{x}) = \int \frac{\mathrm{d}^{3}\mathbf{p}}{(2\pi)^{3}}\left(\zeta^{\,}_{\mathbf{p}}(t)\,a^{\,}_{\mathbf{p}} + \zeta^{*}_{-\mathbf{p}}(t)\,a^{\dagger}_{-\mathbf{p}}\right)\,e^{i\mathbf{p}\cdot\mathbf{x}}\,.
\end{equation}
Here $\zeta_{\mathbf{p}}(t)$ is by definition the solution of equation
(\ref{Mukh}) or (\ref{EoM}) in Fourier space. Therefore any
interaction term proportional to (\ref{EoM}) will necessarily
vanish and give no contribution because we are expanding in terms of
the solutions to that equation. 

Since $\zeta$ on super-horizon scales converges at late times we
should expect both power spectra and bispectra to converge too. This
is not obvious from the form of the action (\ref{action}) as it
requires all terms in $S_{3}$ to converge fast enough at late
times. After horizon crossing $\dot{\zeta} \propto a^{-2}$ therefore
the $a^{3}\zeta\dot{\zeta}^{2}$ terms in $S_{3}$ decay like $a^{-3}$
and $a^{-1}$ at late times respectively. The same is true for the
terms involving $\partial^{-2}\dot{\zeta}$. The
$a\zeta(\partial\zeta)^{2} \to ak^{2}\zeta_{k}^{3}$ term grows like
$a$ at late times however. This appears problematic but it will turn out
that this divergence gives no contribution to \fnl\ and will ultimately
be discarded.

The final term $\propto a^{3}\dot{\zeta}\zeta^{2}$ is problematic. It
grows like $a$ at late times and unlike the $a\zeta(\partial\zeta)^{2}$ term we
are not be able to disregard it. One may neglect this term if one
assumes certain certain conditions\footnote{For
  example if $\epsilon - \eta$ is sufficiently constant as assumed in
  analytical approximations or if it decays rapidly enough at late
  times as done in \cite{Chen:2006xjb, Chen:2008wn}.} on $\epsilon - \eta$  but this goes
against the spirit of the HSR approach.

The HSR approach also requires a more thorough treatment of boundary
terms that have previously been assumed to vanish. Several total derivatives
arise from integration by parts during the derivation of the action in
the form of
(\ref{action}) and while all the total spatial derivatives can be safely
ignored, one total time derivative may give a non-vanishing
contribution \cite{Arroja:2011yj}. The contribution, in terms of HSR parameters, is
\begin{equation}
- \int \mathrm{d}^{4}x\,\frac{\mathrm{d}}{\mathrm{d}t}\left[(\epsilon - \eta)\epsilon a^{3}\zeta^{2}\dot{\zeta}\right]\,,
\end{equation} 

\begin{figure}[t]
  \begin{center}
    \includegraphics[width=8.5cm,trim=0cm 0cm 0cm 0cm,clip]{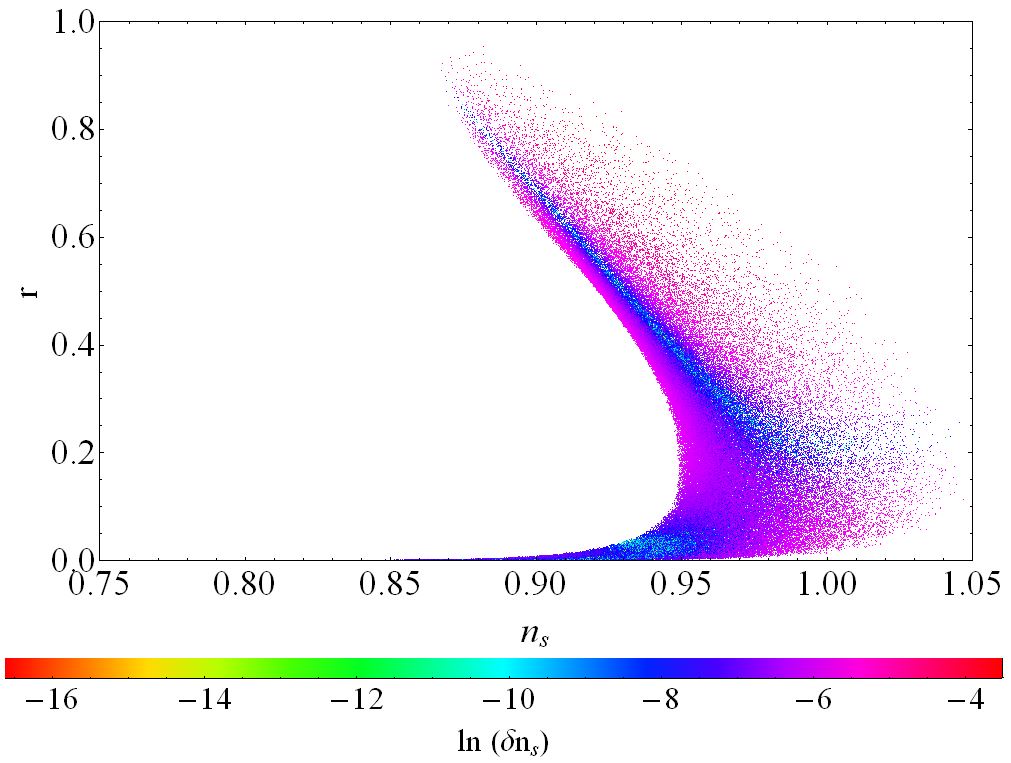} 
    \caption{$r$ {\sl vs} $n_{s}$ scatter plot for $10^{5}$
      trajectories generated as part of the HSR ensemble. Colour
      represents relative difference from the second order slow-roll
      formula for $n_{s}$. $k_{*} = 10^{-5} (\text{Mpc})^{-1} $. The
      distribution clearly shows the typical inflationary
      ``attractor'' for  trajectories with $r > 0$.}
    \label{fig:ns_r}
  \end{center}
\end{figure}

Noting the similarity between the boundary term, the apparently
divergent $a^{3}\zeta^{2}\dot{\zeta}$ term, and the first term in $f(\zeta)$, we write the final line in (\ref{action}) as
\begin{eqnarray}
\int \mathrm{d}^{4}x
\left[a^{3}\epsilon\frac{\mathrm{d}t}{\mathrm{d}t}\left(\epsilon
      - \eta\right)\zeta^{2}\dot{\zeta} + a(\epsilon
    - \eta)\frac{\delta L}{\delta \zeta} -\right. \nonumber\\
\left.\frac{\mathrm{d}}{\mathrm{d}t}\left(a^{3}\epsilon(\epsilon - \eta)\right) + f'(\zeta)\frac{\delta L}{\delta \zeta}\right]\,.
\end{eqnarray}
Here the function $f'(\zeta)$ contains only derivatives of $\zeta$. It
is then straightforward to verify that several cancellations
occur in the first three terms resulting in
\begin{equation}
-2a^{3}\epsilon(\epsilon - \eta)\zeta\dot{\zeta}^{2} - a\epsilon(\epsilon - \eta)\zeta^{2}\partial^{2}\zeta\,.
\end{equation}
The divergent $\zeta^{2}\dot{\zeta}$ disappears in exchange of
$\zeta^{2}\partial^{2}\zeta$ which can be dealt with in the same
manner as the $\zeta(\partial\zeta)^{2}$ term as described
below\footnote{Note also that the remaining terms proportional to the
  equation of motion contain only derivatives of $\zeta$ and can be
  disregarded exactly at the boundary (late times) in the approach
  taken by \cite{Maldacena:2002vr}.} We can then
finally write the action as
\begin{eqnarray}\label{action_final}
  S_{3} &=& \!\!\int d^4x\, a^{3}\epsilon \left[\left(2\eta - \epsilon\right)\zeta \dot{\zeta}^{2} + \frac{1}{a^{2}}\epsilon\zeta(\partial\zeta)^{2}\right.\nonumber\\
  & &\!\!\!\!\left.  - (\epsilon - \eta)\zeta^{2}\partial^{2}\zeta - 2\epsilon\left(1 - \frac{\epsilon}{4}\right)\dot{\zeta}\partial_{i}\zeta\partial_{i}\partial^{-2}\dot{\zeta}\right.\nonumber\\
  & & \!\!\!\!\left. + \frac{\epsilon^{2}}{4}\partial^{2}\zeta\partial_{i}\partial^{-2}\dot{\zeta}\partial_{i}\partial^{-2}\dot{\zeta}\right]\,,
\end{eqnarray}
where we have dropped terms proportional to the first order equation
of motion.

\subsubsection{Numerical Calculation of \fnl}

Using (\ref{action_final}) to define the interaction Hamiltonian one
can use equations (\ref{in_in}) and (\ref{zeta_quantised}) to
calculate the bispectrum. It can be written in the general form
\begin{equation}\label{bispectrum}
B(k_{1}, k_{2}, k_{3}) = {\cal I}\left[\zeta^{*}_{1}\zeta^{*}_{2}\zeta^{*}_{3}\int_{N_{0}}^{N_{2}} dN\, Z(N)\right]\,,
\end{equation}
where ${\cal I}[z]$ distinguishes the imaginary part of $z$, $N_{2}$
and $N_{0}$ are defined $e$-folds (times) defined such that all modes
are deep inside and far outside the horizon respectively (using the
previously defined $A$ and $B$ parameters), $\zeta_{i} =
\zeta_{k_{i}}$. There is a
contribution to $Z(N)$ for each term in the action. For example, the
$\zeta (\partial\zeta)^{2}$ and $\zeta^{2}\partial^{2}\zeta$ terms
give the following contribution
\begin{eqnarray}\label{k2_terms}
\frac{10}{3H}\left[a\epsilon^{2}(\mathbf{k}_{1}\cdot\mathbf{k}_{2}+\mathbf{k}_{1}\cdot\mathbf{k}_{3}+\mathbf{k}_{2}\cdot\mathbf{k}_{3})
  + \right.\nonumber\\
\left.a\epsilon(\eta - \epsilon)(k_{1}^{2}+k_{2}^{2}+k_{3}^{2})\right]\zeta_{1}\zeta_{2}\zeta_{3}\,.
\end{eqnarray}
From (\ref{action_final}), these are the only terms which do not
obviously converge. However, we know at late times $\zeta_{k}\to A_{k}
+ \frac{B_{k}}{a^{2}}$ for some $k$-dependant constants. Considering
the case $k=k_1=k_2=k_3$ for simplicity
\begin{equation}
\zeta^{*3}_{k}\int dN\, a \zeta^{3}_{k} \approx |A|^{6} \int dN\, a + \dots\,,
\end{equation}
where $\dots$ denote terms that converge at late times like
$a^{-1}$. Only the real part of this expression diverges and we are
only interested in the imaginary part for the bispectrum. Therefore
these terms cause no issues at late times, unlike the
$a^{3}\zeta^{2}\dot{\zeta}$ term.

We now specialise to the case where  $k_{1} = k_{2} =
k$ and $k_{3} = \beta k$. This allows us to parametrise most shapes of
interest via the parameter $\beta$ separately from the overall scale
dependence given by wavenumber $k$. Squeezed, equilateral
and folded limits correspond to $\beta = 0$, 1 and 2
respectively. In terms of this classification we can
write down our full expression for \fnl\ as
\begin{eqnarray}\label{fnl_prelim}
\!\!\!\!\!\!\!f_{\mathrm{NL}} &=& \frac{1}{|\zeta|^{2}\left(|\zeta|^{2} +
    2|\zeta_{\beta}|^{2}\right)}\times\nonumber\\
&&\!\!\!\!\!\!\!\!\!\!\!\!{\cal I}\left[\zeta^{*2}\zeta^{*}_{\beta}\int_{N_{0}}^{N_{2}} dN\, f_{1}\zeta^{2}\zeta_{\beta} + f_{2}\zeta\zeta^{\prime}\zeta_{\beta}^{\prime} + f_{3}\zeta_{\beta}\zeta^{\prime 2}\right]\,,
\end{eqnarray}
where $\zeta = \zeta_{k}, \zeta_{\beta} = \zeta_{\beta k}$ and $\zeta^{\prime} = \mathrm{d}\zeta/\mathrm{d}N$. The functions $f_{i}$ are given by
\begin{eqnarray}
\!\!\!\!\!f_{1} & = & \frac{5k^{2}a\epsilon}{3H}(2 + \beta^{2})(2\eta - 3\epsilon)\,,\nonumber\\
\!\!\!\!\!f_{2} & = & -\frac{10Ha^{3}\epsilon}{3}\left[4\eta + (1-\beta^{2})\epsilon + \left(\frac{\beta^{2}}{4} - 1\right)\epsilon^{2}\right]\,,\\
\!\!\!\!\!f_{3} & = & -\frac{5Ha^{3}\epsilon}{3}\left[4\eta + 2(\beta^{2}-1)\epsilon + \left(\frac{\beta^{2}}{4} - 1\right)\beta^{2}\epsilon^{2}\right]\nonumber\,.
\end{eqnarray}

The last remaining difficulty lies with the early time behaviour of
the integrand. At very early times ($N_{0}\to -\infty, a \to 0, A\to
\infty$) $\zeta$ oscillates very rapidly and has a growing amplitude,
but the \fnl integral formally converges. At early times the integrand
becomes proportional to
\begin{equation}
\int_{-\infty}^{N} dN\, f(H, \epsilon, \dots)\left(\frac{k}{aH}\right)^{n}e^{-i(2 + \beta)\frac{k}{aH}}\,,
\end{equation}
for some integer $n$. By rotating slightly into the imaginary plane,
$(k/aH) \to (1 - i\delta)(k/aH)$ one can obtain a finite answer
independent of the cut-off time. Numerically one cannot integrate to
infinity and in it's present form the integral does not converge
numerically. To resolve this one can add a damping factor to the integrand (similar
to the above procedure) however this tends to systematically
underestimate the final integrals and the optimum damping factor
$\delta$ differs from mode to mode \cite{Chen:2006xjb,Chen:2008wn}.

A better method is to use the early time approximation for $\zeta$ and
then integrate by parts. We are interested in calculating an integral
of the form
\begin{equation}\label{eq:int}
I = \int_{-\infty}^{N} dN\, f(N)\,\zeta^{2}\zeta_{\beta}\,.
\end{equation}
Using (\ref{Initial_zeta}) we can write $\zeta^{2}\zeta_{\beta}$ at
early times as
\begin{equation}\label{early_approx}
\zeta^{2}\zeta_{\beta}  \to  \frac{1}{\Gamma}\frac{\mathrm{d}}{\mathrm{d}N}(\zeta^{2}\zeta_{\beta})\,,
\end{equation}
where 
\begin{equation}
\Gamma =  -\left[i(2 + \beta)\frac{k}{aH} + 3(1 + \epsilon - \eta)\right]\,.
\end{equation}

Inserting this into (\ref{eq:int}) and integrating by parts yields
\begin{equation}
I \to \left[\frac{f(N)}{\Gamma}\zeta^{2}\zeta_{\beta}\right]_{-\infty}^{N} - \int_{-\infty}^{N} dN\,\frac{\mathrm{d}}{\mathrm{d}N}\left(\frac{f(N)}{\Gamma}\right)\zeta^{2}\zeta_{\beta}\,.
\end{equation}

The resulting integral is now more convergent than before as $1/\Gamma
\to aH/k$. One can repeat the process until the final integrand
converges in the limit $a \to 0$ and all divergences are transferred
to the boundary term. These divergences can be removed by using the
same contour as before, but now the terms vanish for any finite
$\delta$. The boundary term evaluated at $N = -\infty$ can then be
safely ignored.

To apply this procedure to the calculation of \fnl we first split the integral into two parts
\begin{equation}\label{eq:split}
\int_{N_{0}}^{N_{2}} dN\, = \int_{N_{0}}^{N_{1}} dN\, + \int_{N_{1}}^{N_{2}} dN\,,
\end{equation}
where $N_{0}$ and $N_{2}$ are times when $k = A\,aH$ and $k = B\,aH$
respectively with $A \gg 1$ and $B \ll 1$. $N_{1}$ is any time when
(\ref{Initial_zeta}) is a good approximation for both
modes. The late time contribution remains unchanged and we perform the
``approximate then integrate by parts'' procedure to the early time
contribution. The early time contribution, $E$, then takes the form
\begin{eqnarray}\label{early_contribution}
E &=& \left.\frac{5Ha^{3}\epsilon}{3(2+\beta)^{3}}\left[B_{1}\Gamma +
    \dots +
    \frac{B_{-4}}{\Gamma^{4}}\right]\zeta^{2}\zeta_{\beta}\right|_{N_{1}}
-\nonumber\\
&&\!\!\!\!\!\!\!\!\!\!\!\! \int_{N_{0}}^{N_{1}} dN\, \frac{5Ha^{3}\epsilon}{12(2+\beta)^{3}}  \left[\frac{A_{-2}}{\Gamma^{2}} + \dots + \frac{A_{-6}}{\Gamma^{6}}\right]\zeta^{2}\zeta_{\beta}\,,
\end{eqnarray}
where $A_{i}$ and $B_{i}$ are polynomials of the HSR parameters and $\beta$. For example 
\begin{eqnarray}
B_{1} &=& (2 + \beta)^{2}\left[\left(4 + \beta(2\beta -
    3)\right)\epsilon - \right. \nonumber\\
  &&\left.2(2+\beta)\eta +\beta\left(1-\frac{\beta^{2}}{4}\right)\epsilon^{2}\right]\,.
\end{eqnarray}
We omit the full list of the complicated polynomials for brevity. The second
term in (\ref{early_contribution}) gives a completely
negligible contribution to the final value of \fnl\ as it is
roughly a factor of $\Gamma^{3}$ smaller and we are in the regime
where $\Gamma >> 1$. The early time contribution is therefore given
completely by the boundary term in (\ref{early_contribution}).

This method was first used in \cite{Chen:2008wn}. However the authors
choose to focus on particular inflation models such as those with a
feature whereas this paper takes a much more general approach. Dealing
with the late time divergence from $\zeta^{2}\zeta^{\prime}$ also
received little attention. The best explanation on how to deal with
this is in \cite{Hazra:2012yn} where the authors demonstrate a
fortunate cancellation between the troublesome term and the field
redefinition.

Here we explicitly keep all terms to all orders in slow-roll. Most of
the computational effort is spent dealing with the oscillatory nature
of $\zeta$ so not much is gained by a slow-roll approximation. This
allows a much broader range of models to be analysed which in turn
leads to Monte Carlo treatment in the next section. We do drop the
early time integration in (\ref{early_contribution}) but this is an
approximation relying on the behaviour of $\zeta$ in the limit $k \gg
aH$, not an explicit slow-roll approximation. Finally, to our
knowledge, this is the first time the third order action has been
presented in the form of (\ref{action_final}) and used in a
calculation. This form provides a much more efficient way to perform
the numerical calculation without having to rely on fortuitous
cancellations of terms after the integration.

In summary \fnl, to a good approximation with respect to the early
time oscillatory integral, is given by the following expression
\begin{eqnarray}\label{fnl_final}
  f_{\mathrm{NL}} & =& \left[\left.|\zeta|^{2}\left(|\zeta|^{2} +
        2|\zeta_{\beta}|^{2}\right)\right|_{N_{2}}\right]^{-1}\times\\
  &&\!\!\!\!\!\!\!{\cal I}\left[\left.\zeta^{*2}\zeta^{*}_{\beta}\right|_{N_{2}}\int_{N_{1}}^{N_{2}} dN\, \left(f_{1}\zeta^{2}\zeta_{\beta} + f_{2}\zeta\zeta^{\prime}\zeta_{\beta}^{\prime} + f_{3}\zeta_{\beta}\zeta^{\prime 2}\right)\right.\nonumber\\
  &&\!\!\!\!\!\!\! \left. +
    \left.\zeta^{*2}\zeta^{*}_{\beta}\right|_{N_{2}}\left.\frac{5Ha^{3}\epsilon}{3(2+\beta)^{3}}\left[B_{1}\Gamma
        + \dots +
        \frac{B_{-4}}{\Gamma^{4}}\right]\zeta^{2}\zeta_{\beta}\right|_{N_{1}}\right] \,.\nonumber
\end{eqnarray}

\section{Results}\label{results}

\begin{figure*}[t]
  \begin{center}
    \begin{tabular}{cc}
      \makebox[8.5cm][c]{
    \includegraphics[width=8.5cm,trim=0cm 0cm 0cm
    0cm,clip]{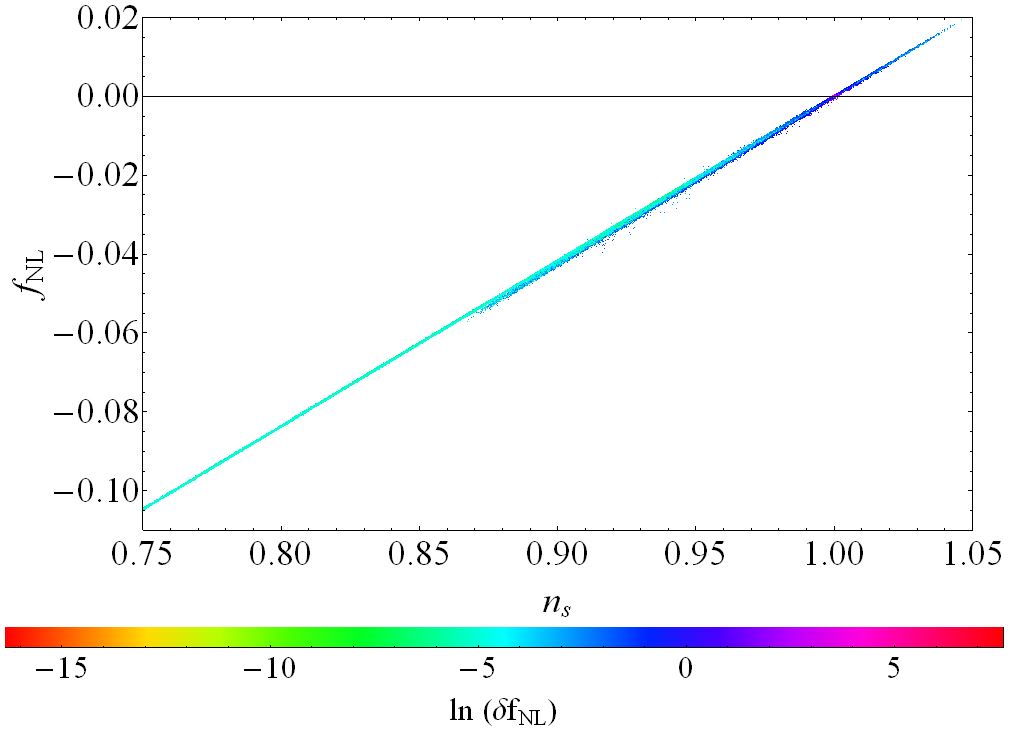}}&
  \makebox[8.5cm][c]{
    \includegraphics[width=8.5cm,trim=0cm 0cm 0cm 0cm,clip]{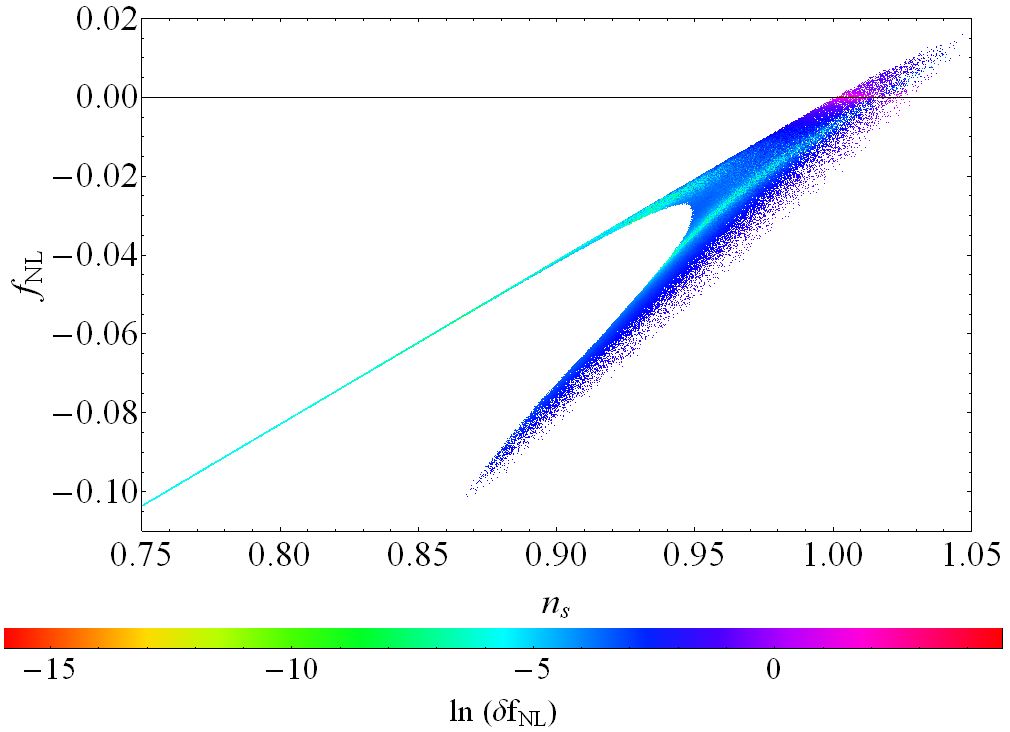} 
  }
  \end{tabular}
    \caption{$n_{s}$ {\sl vs} \fnl\  scatter plot for $10^{5}$
      trajectories generated with ``end-of-inflation'' priors. The
      left panel is for the squeezed limit $\beta=0.1$ and the right
      panel is for the equilateral case $\beta=1$. The colour scale
      represents the $\ln$ of the relative difference from the
      slow-roll approximation for \fnl. The values of $n_s$ and \fnl\
      are sampled for a scale corresponding to $k_\star = 10^{-5} (\text{Mpc})^{-1} $.}
    \label{fig:ns_fnl_squeezed}
  \end{center}
\end{figure*}

As a check of our method we have verified that our results converge on
super-horizon scales and with respect to early-time integration
limits. The first condition is illustrated in
Figure~\ref{fig:time_dependance} for a typical random trajectory drawn
from the ensemble generated by our method using the end-of-inflation
random boundary conditions on the HSRs. A typical trajectory in these
ensembles will be deep in the slow-roll regime when modes of interest
cross the horizon. The green line is the real part of
$\zeta$ while the red and blue lines represent the real and imaginary
parts of \fnl as a function of $N$. \fnl\ oscillates roughly three
times quicker than $\zeta$ as it is proportional to $\zeta^{3}$. The
real part diverges due to the $k^{2}a\zeta^{3}/H$ term discussed
previously however it does not contribute to the amplitude of the
correlator in the in-in formalism and can be safely ignored. The
imaginary part (the value of interest) converges when the mode
exits the horizon. The results shown in
figure~\ref{fig:time_dependance} does not include the contribution of
the boundary term in (\ref{fnl_final}) as it contributes only a
constant.

The next step is to verify our results do not depend sensitively on
the early time cut-off. Figure~\ref{fig:early_time} shows the dependence
of \fnl\ as the integration is started at earlier and earlier
times. The color represents the value of $\beta$, our shape parameter
for the ${\mathbf k}_1$+${\mathbf k}_2$+${\mathbf k}_3$ triangle. The
value of \fnl\ converges for all shapes when the parameter $A$, which
sets how much smaller than the horizon the mode with the {\sl
  smallest} $k$ in the triangle ${\mathbf k}_1$+${\mathbf
  k}_2$+${\mathbf k}_3$ has to be at the start
of integration, is approximately 400. This is larger than what would
be required for an accurate calculation of the corresponding power
spectrum statistic due to the diverging oscillatory behaviour of the
terms contributing to the \fnl\ integration.

It is also important to verify convergence with respect to the choice
of integration split point $N_{1}$, or cut-off time, introduced in
(\ref{eq:split}).The choice is parametrised by the variable  $X$ defined by $X =
k/aH$, again this condition is imposed on the {\sl
  smallest} $k$ in the triangle ${\mathbf k}_1$+${\mathbf
  k}_2$+${\mathbf k}_3$. \fnl\ as a function of $X$ is shown in
figure~\ref{fig:cutoff_point}. If $X$ is too small, the split point is too
close to the time of horizon exit and the early time approximation used in
(\ref{early_approx}) will not be valid. If $X \sim A \to \infty$, this is
equivalent to (\ref{fnl_prelim}) i.e. doing no regularisation
procedure at all. Therefore if $X$ is too large relative to $A$ one
would expect the early time contribution to be unable to compensate
for the increasingly divergent integral. This is the origin of the
noise seen in figure~\ref{fig:cutoff_point}. There is an optimal region
for the value of
$X$ which minimises the combined contribution from both sources of
numerical 
error. From figure~\ref{fig:cutoff_point} it can be seen that $\ln X
\approx 4 - 5$ is a good choice for ``folded'' shapes $\beta\to 2$
(left-panel). There optimal position for the split-point is somewhat
shape dependent as shown in the right-panel of
figure~\ref{fig:cutoff_point} which shows ten ``squeezed'' cases for
the same HSR trajectory but in both cases for $\ln \sim 4$ the
inaccuracies are very small ($\ll 1$\%). For the following we chose
the values $\ln A=6$, $\ln X=5$, and $B$, the parameter that sets the
required size of the largest $k$ in the ${\mathbf k}_1$+${\mathbf
  k}_2$+${\mathbf k}_3$ triangle with respect to the horizon at the
end of the integration, is set to 0.01.

We generate ensembles of trajectories for two different HSR
  boundary conditions. The first is the ``end-of-inflation'' setup
where the HSR are drawn from uniform distributions with a given range
at the end of inflation defined by the time when $\epsilon=1$. The
second, ``early-time'' case is one where the HSR, including $\epsilon$
in this case, are drawn from uniform distributions at the time when
the largest scale of interest is crossing the horizon. For this case
$\epsilon$ is drawn from the range $[0,0.4]$ and the system is evolved
{\sl back} $\ln A=6$ $e$-folds to the start of the mode integration
and then forward for the required number of total $e$-folds to cover
horizon exit of all observables scales. 

For both cases we used $l_{\text{max}} = 4$ and $s = 1.5$ as defined
in (\ref{uniform}) to impose a hierarchical prior. For the
``end-of-inflation'' ensemble this choice is wide enough to give a
proposal distributions in the observables $n_s$, $r$, etc. that are
wider than the current, parametric constraints obtained from the
recent Planck analysis \cite{Ade:2013uln}. For each trajectory the
number of $e-$folds was chosen from a uniform distribution in the range
be $N_{\text{tot}} = [60,80] + \ln A$. The factor of $\ln A$ is
important to maintain convergence in the limit of $A \to \infty$ as
discussed previously. Each ensemble includes some ${\cal O}(10^5)$
trajectories.

In figure~\ref{fig:shape_scale_dependence} we show \fnl\ as a function
of shape parameter $\beta$ and overall scale $k$ for a selection 30
trajectories from the ``end-of-inflation'' ensemble. For this ensemble
we expect that at the time when observable quantities are evaluated
the HSRs are going to be in the deep slow-roll limit with
$^{i}\lambda\ll 1$. This is due to the fact that the system is evolved
back from the wide proposal at the end of inflation towards a
slow-roll attractor at early times when the observable scales are
exiting the horizon. The results for this ensemble should therefore
agree with the slow-roll approximations and consistency
conditions. Figure~\ref{fig:shape_scale_dependence} shows that the
scale dependence is very mild and that for trajectories where there is
shape dependence $|$\fnl$|$ peaks close to the equilateral configuration
$\beta=1$. It is also known that \fnl\
should be near scale-invariant in the slow-roll limit and peak in the
equilateral configuration. In addition, \fnl\ must also satisfy the
well known consistency condition in the squeezed limit given by
$f_{\mathrm{NL}}\approx (5/12)(n_{s} - 1)$
\cite{Maldacena:2002vr,Creminelli:2004yq}.

As a consistency check we also make scatter plots for the ensembles in
the $n_s$ {\sl vs} $r$ and $n_s$ {\sl vs} \fnl\ planes. We do this by
plotting the values of $n_s$, $r$, and \fnl\ from the {\sl largest}
scale for each trajectory in the ensembles. In the slow-roll limit the
$n_s$ {\sl vs} $r$ plane should show a clear ``inflationary''
attractor \cite{Liddle:1994dx, PhysRevD.42.3936}. The \fnl\ consistency condition
should also appear as a strong attractor in the squeezed $\beta\sim 0$
shape case. 

Figure~\ref{fig:ns_r} shows the ``end-of-inflation'' ensemble scatter
plot for $n_s$ {\sl vs} $r$. The inflationary attractor is clearly
visible. The colour coding in the figure depicts the difference
between the numerical $n_s$ and second order slow-roll approximation
$\bar n_s$ given by (\ref{nsr_eqn1}) and defined $\delta n_{s} =
\left|(n_{s} - \bar{n}_{s})/n_{s}\right|$. This shows that the
numerical and slow-roll results for $n_{s}$ agree very well when the
trajectory lies close to the attractor. 

\begin{figure}[t]
  \begin{center}
    \includegraphics[width=9cm,trim=1cm 5cm 1cm 2cm,clip]{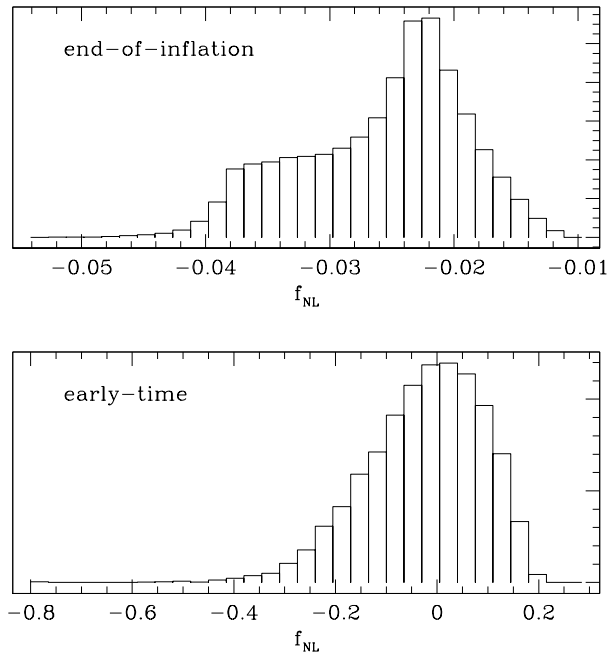} 
    \caption{Histogram of \fnl\ values equilateral bispectra for the
      large scale mode $k_\star = 10^{-5} (\text{Mpc})^{-1}$ in both
      ``end-of-inflation'' (top) and ``early-time'' (bottom)
      ensembles. Both ensembles have been filtered such that all
      trajectories have $0.946 < n_s < 0.976$ at the {\sl smaller} scale
      $k = 10^{-2} (\text{Mpc})^{-1}$ in order to agree roughly with
      observations at the 2$\sigma$ level. The ``early-time'' proposal
      of HSR parameters allows for significant variation in the
      parameters while the largest scales are crossing the horizon
      leading to \fnl\ about an order of magnitude larger than in the
      other case.}
    \label{fig:prior_2}
  \end{center}
\end{figure}

The equivalent of the slow-roll expressions (\ref{nsr_eqn1})-(\ref{nsr_eqn3}) for \fnl\ is
\begin{equation}\label{fnl_approx}
\bar{f}_{\mathrm{NL}} = \frac{5}{12}\left(\bar{n}_{s} - 1 + f(\beta)\,\bar{n}_{t}\right)\,,
\end{equation}
where $\bar n_t$ is the slow-roll approximation for the tensor
spectral index and $f(\beta)$ is a function of the shape with
$f(\beta) \to 0$ as $\beta \to 0$ and $f(\beta) = 5/6$ when $\beta =
1$. Even though this formula was derived only at first order in
$\epsilon$, $\eta$ we used the second order formulae for $n_{s}$ and
$n_{t}$. Figure~\ref{fig:ns_fnl_squeezed} shows 
the trajectories in the $n_{s}$ {\sl vs} \fnl\ plane for both the squeezed
and equilateral. The $5/12\, n_s$ dependence is clear in both cases
but the equilateral case has an additional dependence on $n_t$ which
dominates when $n_s\to 1$ in analogy with
Figure~\ref{fig:ns_r}. The figure also shows the difference between
the slow-roll approximation for \fnl\ and the value obtained
numerically. The two agree to within a few percent except when
\fnl$\ll 10^{-2}$.

Figure~\ref{fig:prior_2} shows what happens to the equilateral \fnl\
distributions in the case where the trajectories are generated using
the ``early-time'' priors on the HSR parameters. In this case, if the
proposal ranges for the HSR are wide enough, the largest scales
considered will be crossing the horizon when the trajectory is
typically still in the out-of-slow-roll regime. At later times the
trajectory will typically end up in a slow-roll attractor and the
situation will revert to a picture much closer to that seen in
figure~\ref{fig:ns_fnl_squeezed}. The squeezed distribution remains
unchanged but the equilateral case can have \fnl\ values much larger
than that allowed by the $5/12\, n_s$ scaling. Typically the value of
$n_s$ for the scale where we are sampling \fnl\ is also large but we
have filtered the trajectories to include only ones where $0.946 < n_s
< 0.976$ at the {\sl smaller} scale $k = 10^{-2} (\text{Mpc})^{-1}$
where observational constraints are much tighter. The filter imposes a
severe cut on the trajectories with only a fraction $\sim 10^{-3}$ of
trajectories satisfying the constraint on $n_s$ on smaller scales. For
this subset of trajectories the power spectrum, on the largest scales,
has a strong scale dependence. This may be preferred by observations
of the CMB where there are indications of lower than expected power on
the largest scales.

\section{Discussion}\label{conclusion}

We have outlined a full numerical calculation of the bispectrum of
primordial curvature perturbations arising from generalised
inflationary trajectories. The bispectrum has been evaluated in terms
of a scale dependent \fnl$({\mathbf k}_1,{\mathbf k}_2,{\mathbf
  k}_3)$. The calculation is valid in the out-of-slow-roll regime as
long as the weak coupling limit is maintained. This is of interest in
models where there is significant evolution of slow-roll parameters
during inflation that can lead to observational features in both power
spectrum and bispectrum.

We have explored the generation of inflationary ensembles via the HJ
formalism using HSR parameters and calculated the distribution of the
bispectrum \fnl\ for various configurations of the ${\mathbf
  k}_1+{\mathbf k}_2+{\mathbf k}_3$ triangle. In doing so we have
verified the consistency relation for the squeezed limit and
the equilateral configurations in the slow-roll regime. We have shown
that, in the out-of-slow-roll limit, \fnl\ equilateral has a much
wider distribution due to the scale dependence of the perturbations
and has values that are typically an order of magnitude larger than in
the slow-roll limit. These types of trajectories can be viable with
respect to observations since on smaller scales the perturbations
become near scale invariant due to the HSR asymptoting to small
values.

The generation of inflationary ensembles including the calculation of
the bispectrum will be useful for HSR parameter explorations using
future data. \fnl\ observational constraints are currently far from
the regime where they can affect the shape of trajectories and
consequently add to our knowledge of the shape of the inflaton
potential. However future observations may probe a regime that could
constrain any out-of-slow-roll features in the trajectories. This
would in turn constrain any significant feature in the single field
inflation scenario. Even if features do not exist, probing \fnl\ to
${\cal O}(10^{-2})$ by a combination of future LSS observations would
be a powerful probe of inflationary physics, particularly in scenarios
where no tensor perturbations are detected. 

\begin{acknowledgements}
  JSH is supported by a STFC studentship. CRC and JSH acknowledge the
  hospitality of the Perimeter Institute for Theoretical Physics  and
  the Canadian Institute for Theoretical Astrophysics where
  some of this work was carried out.
\end{acknowledgements}

\bibliography{paper}

\end{document}